\documentclass[%
reprint,
groupedaddress,
amsmath,amssymb,
aps,
prl,
longbibliography,
superscriptaddress]{revtex4-1}
\usepackage{graphicx}
\usepackage{dcolumn}
\usepackage{bm}
\usepackage{physics}
\usepackage{color}
\RequirePackage[colorlinks=true,linkcolor=blue,urlcolor=blue,hyperfootnotes=false,citecolor=blue]{hyperref}

\newcommand\at[2]{\left.#1\right|_{#2}}

\begin{document}

\title{Demonstration of Weak-Link Physics in the Dynamical Response of Transition-Edge Sensors}

\author{Marios Kounalakis}
 \email{marios.kounalakis@gmail.com}
\affiliation{Kavli Institute of Nanoscience, Delft University of Technology, 2628 CJ Delft, The Netherlands}

\author{Luciano Gottardi}
\affiliation{NWO-I/SRON Netherlands Institute for Space Research, Niels Bohrweg 4, 2333 CA Leiden, Netherlands}

\author{Martin de Wit}
\affiliation{NWO-I/SRON Netherlands Institute for Space Research, Niels Bohrweg 4, 2333 CA Leiden, Netherlands}

\author{Yaroslav M. Blanter}
\affiliation{Kavli Institute of Nanoscience, Delft University of Technology, 2628 CJ Delft, The Netherlands}
\date{\today}

\begin{abstract}
We theoretically predict and experimentally observe the onset of weak-link physics in the dynamical response of transition edge sensors (TES).
We develop a theoretical framework based on a Fokker-Planck description that incorporates both the TES electrical response, stemming from Josephson phenomena, and the electrothermal effects due to coupling to a thermal bath.
Our measurements of a varying dynamic resistance are in excellent agreement with our theory, thereby establishing weak-link phenomena as the main mechanism underlying the operation of TES.
Furthermore, our description enables the calculation of power spectral densities, paving the way for a more thorough investigation of the unexplained ``excess noise'' in long diffusive junctions and TES reported in recent experiments.
\end{abstract}
\maketitle

\section{Introduction}

Superconducting transition-edge sensors (TES) are extremely sensitive thermometers~\cite{moseley1984thermal, ullom2015review} used as microcalorimeters and bolometers in ground- and space-borne low-temperature instruments for the detection of radiation from millimeter-wave up to gamma rays in the fields of astrophysics~\cite{spilker2018fast,miller2018massive,carlstrom201110}, plasma physics~\cite{eckart2021microcalorimeter}, particle physics~\cite{alpert2015holmes} and nuclear material analysis~\cite{wollman1997high,hoover2009microcalorimeter}, as well as quantum information~\cite{shen2018randomness}.
Moreover, their high energy resolution in the X-ray range has granted their selection as primary detectors in large-scale telescope missions, such as ESA's forthcoming \emph{Athena} mission~\cite{nandra2013hot,ravera2014x,barret2018athena}.
These low-impedance devices are made of a thin superconducting bilayer and are electrically biased close to their critical temperature, $T_c$, via superconducting leads at constant (dc) or alternating (ac) current, depending on the read-out scheme~\cite{ullom2015review,gottardi2021review}.
During a detection event, the TES temperature rises proportionally to the photon energy, leading to a sharp change in the TES resistance as a result of the sharp superconducting-to-normal transition~\cite{irwin2005transition}.
Understanding the underlying physical mechanism during this transition is, therefore, crucial for predicting the detector sensitivity.

Over the past decade there has been rapid progress in understanding the physics of TESs~\cite{sadleir2010longitudinal,bennett2013resistance,gottardi2014josephson,gottardi2021voltage}, however, their noise properties are not yet fully understood and there is currently no consensus on the model describing the physical mechanism underlying the TES resistive transition and dynamical response~\cite{bennett2013resistance,wessels2021model}.
The theoretical models typically employed are based either on weak-link physics, \emph{i.e.}, using the Resistively Shunted Junction (RSJ) model~\cite{kozorezov2011modelling}, or on physical mechanisms occurring in the film, such as the two-fluid model~\cite{irwin1998thermal} and the less investigated Berezinskii--Kosterlitz--Thouless (BTK) theory~\cite{fabrega2018large}.
Despite showing good agreement with the experimental data at equilibrium, the models are typically tailored to fitting existing data in a phenomenological fashion, lacking the predictive power of a full theoretical description.
As a result, a universal description of the underlying physical mechanism involved in the TES dynamical response, which determines the detector sensitivity to current and temperature, is still missing.

Here we develop a theoretical framework, based on weak-link physics, that models the TES electrical response including the electrothermal effects resulting from its coupling to a thermal bath.
More specifically, we employ a Fokker-Planck equation that enables the integration of Josephson physics in the electrically biased circuit with the thermal equilibrium steady state.
Inspired by the large separation between the electrical and thermal response timescales, naturally occurring in dynamical response measurements, we derive an alternative definition of the dynamic resistance and noise sensitivity parameters.
We validate our model predictions against data measured in several TES devices, showing excellent agreement with our theory as opposed to existing models.
In particular, we find a significantly varying dynamic resistance across the resistive transition, which is a key signature of Josephson physics.
Our experimental observations in combination with our theoretical predictions, provide clear evidence of weak-link phenomena underlying the physics of TES.
Finally, we show that our model allows for the direct calculation of noise power spectral densities, enabling the investigation of mixed-down noise from high-frequency processes, which is believed to be the origin of unexplained ``excess noise''~\cite{wessels2021model,gottardi2021voltage,dewit2021impact}.

\begin{figure}[t]
  \begin{center}
    \includegraphics[width=1.0\linewidth]{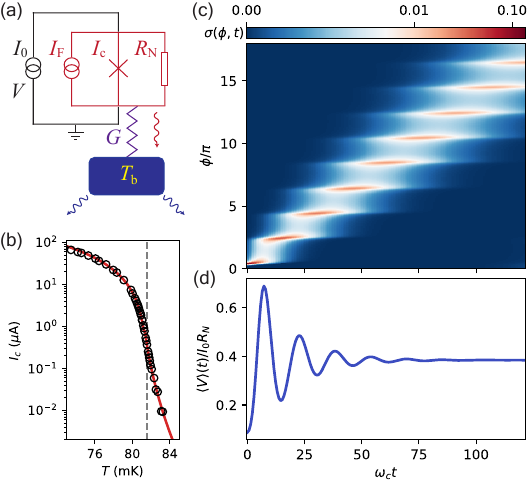}
  \end{center}
  \caption{
  {(a)~Circuit description of TES, which we model as an electrically biased Josephson junction with critical current, $I_c$, in parallel to the normal state resistance, $R_N$, and thermally fluctuating current, $I_F$.
  The TES is additionally coupled to a thermal bath at temperature $T_b$ via the thermal conductance $G$.
  (b)~Plot of $I_c$ as a function of temperature, $T$, obtained by fitting the experimental data (open circles) with a combined Ginzburg-Landau and weak-link theory model, where the dashed line indicates the critical temperature (see text for details).
  (c)~Example of Fokker-Planck dynamics in the electrical circuit (not including the thermal coupling to $T_b$) for $I_0/I_c=1.08$ ($I_c=6.74~\mathrm{\mu A}$) and $\gamma=4000$ showing the evolution of $\sigma(\phi,t)$ according to Eq.~(\ref{eq:Fokker-Planck}) plotted in characteristic time units, where $\omega_c=I_cR_N/\phi_0$.
  (d)~Corresponding average voltage $\langle V\rangle(t)$ normalized by $I_0R_N$.}
  }
  \label{fig:device}
\end{figure}

\section{Theoretical framework}

A typical TES comprises a bilayer thin film that is electrically biased via two superconducting electrodes.
Several experiments have shown that the latter proximitize the bilayer film, resulting in the formation of a ${SS^\prime S}$ weak link.
Supporting evidence includes the observation of Fraunhofer-like oscillations of the critical current $I_c$ upon application of a perpendicular magnetic field~\cite{sadleir2010longitudinal}, the exponential dependence of $I_c$ on the temperature $T$ and TES length $L$~\cite{sadleir2011proximity,ridder2020study}, the scaling of the transition width $\sim1/L^2$, as well as the observation of a non-linear inductance in ac-biased devices~\cite{gottardi2014josephson,gottardi2021voltage}.

These observations suggest that the electrical response of TES is governed by Josephson physics originating in the coupling between the electrodes~\cite{josephson1962possible,tinkham2004introduction}.
For the circuit description of electrically biased junctions one typically uses the resistively shunted junction (RSJ) model~\cite{ambegaokar1969voltage}, which consists of a current-biased Josephson junction in parallel to a normal resistor, $R_N$, in the presence of a fluctuating current $I_F(t)$, as schematically depicted in Fig.~\ref{fig:device}(a).
The latter represents Gaussian white noise at temperature $T$, with $\langle I_F(t)\rangle=0$ and ${\langle I_F(t)I_F(t^\prime)\rangle=\delta(t-t^\prime)2k_BT/R_N}$.
Within this model, the total current flowing through the device is given by $I_0=I_c\sin{\phi}+V(t)/R_N+I_F(t)$, where $I_c$ is the critical current of the junction, $\phi$ is the gauge-invariant phase difference across the electrodes and according to the Josephson relation, $V(t)=\phi_0{d\phi}/{dt}$, where $\phi_0\equiv \hbar/(2e)$ is the reduced flux quantum~\cite{tinkham2004introduction}.

Here, we model the dynamics described above using a 1-dimensional Fokker-Planck (Smoluchowski) equation for the probability density $\sigma(\phi,t)$~\cite{ambegaokar1969voltage}
\begin{equation}
\frac{\partial \sigma}{\partial{t}} = \omega_c\left[\sigma\cos{\phi}-\left(\frac{I}{I_c}-\sin{\phi}\right)\frac{\partial \sigma}{\partial \phi}+\frac{1}{\gamma}\frac{\partial^2 \sigma}{\partial \phi^2}\right],
\label{eq:Fokker-Planck}
\end{equation}
where $\omega_c=I_cR_N/\phi_0$ is the characteristic frequency and $\gamma=\hbar I_c/(2ek_BT)$ is defined as the Josephson to thermal energy ratio~\cite{kozorezov2011modelling}.
Knowledge of $\sigma(\phi,t)$ enables the calculation of the average phase and voltage evolution, $\langle\phi\rangle=\int{(\sigma\phi) d\phi}$ and $\langle V\rangle(t)=\phi_0\frac{d\langle{\phi}\rangle}{dt}$, respectively, with normalization $\int{d\phi \sigma}=1$.
We solve Eq.~(\ref{eq:Fokker-Planck}) using the Python package GEKKO~\cite{beal2018gekko}.
As an example, in Figs.~\ref{fig:device}(c) and \ref{fig:device}(d) we plot the dynamical evolution of $\sigma$ and the corresponding average voltage, respectively, for $I/I_c=1.08$ and $\gamma=4000$.
As expected, the nonlinear evolution of the phase leads to an oscillating average voltage, which would keep oscillating forever at zero temperature ($\gamma\rightarrow\infty$).
At finite temperature, however, these oscillations are suppressed above a certain timescale when the average phase starts to evolve linearly over time.
For times above this timescale ($\tau^\prime$) the system operates at a steady-state average voltage, $V^\prime \equiv  {\langle V\rangle(\tau^\prime)}$.

For a realistic model description of a TES one should also take into account the temperature dependence of the critical current.
A typical $I_c(T)$ behavior is shown in Fig.~\ref{fig:device}(b), where we plot the measured $I_c(T)$ data for one of our devices (open circles), alongside a numerical fit based on a combined Ginzburg-Landau and weak-link model.
The first is described by $I_c(T) = I_{c0}\left|1-T/T_c\right|^{3/2}$ and the latter by $I_c(T) \propto \left|T/T_c-1\right|^{1/2}e^{-L/\xi_0 \left|T/T_c-1\right|^{1/2}}$, where $I_{c0}$ is the zero-temperature critical current, $T_c$ is the critical temperature, $L$ the TES length and $\xi_0$ its zero-temperature coherence length~\cite{ullom2015review,gottardi2014josephson}.
The vertical dashed line in Fig.~\ref{fig:device}(b) corresponds to the critical temperature.
In our model, for each device, the $I_c(T)$ function obtained from fitting the experimental data is used as input in Eq.~(\ref{eq:Fokker-Planck}) such that $\langle V\rangle(t)$ is only a function of $I$ and $T$.

Furthermore, TES are connected to a thermal bath at temperature, $T_b$, as depicted in Fig.~\ref{fig:device}(a).
This thermal coupling results in electrical power loss as a result of the thermal heat flow~\cite{ullom2015review,gottardi2021review} that is typically described by the power-thermal equation~\cite{irwin1995application}
\begin{equation}
C_H\frac{dT}{dt} = I_0V(t) - K(T^n-T_b^n),
\label{eq:PowerThermal}
\end{equation}
where $C_H$ is the heat capacity of the TES and $K=G/(nT^{n-1})$ is a material- and geometry-dependent parameter, defined by the differential thermal conductance $G$ and the thermal exponent $n$ (typically $3\lesssim n\lesssim4$)~\cite{ullom2015review}.
The TES reaches a steady state, at a time $\tau\gtrsim C_H/G\gg\tau^\prime$, when the generated electrical power is balanced by the thermal power lost to the bath.
In this regime the TES is stable, operating at an equilibrium voltage, $V_0=K(T{}^n-T_0^n)/I_0$, and temperature $T_0$.

\section{Modeling the dynamical response}

Crucial to the performance of a TES is its sensitivity to changes in temperature and current.
This is typically quantified using the parameters $\alpha_I=\at{\frac{T_0}{R_0}\frac{\partial R}{\partial T}}{I_0}$ and $\beta_I=\at{\frac{I_0}{R_0}\frac{\partial R}{\partial I}}{T_0}$, which describe the logarithmic temperature and current sensitivities, respectively~\cite{ullom2015review}.
A high $\alpha_I/\beta_I$ ratio is desirable for better detector performance.
Experimentally, these parameters can be extracted directly by measuring the TES response to a vanishingly small electrical signal, that perturbs the TES resistance as $R(T,I)=R_0+\at{\partial{R}/\partial{T}}{I_0}\delta{T}+\at{\partial{R}/\partial{I}}{T_0}\delta{I}$~\cite{taralli2019complex,gottardi2021voltage}.

Importantly, for an accurate extraction of $\alpha_I$ and $\beta_I$, a low-power perturbing signal is injected into the electro-thermal system in order to maintain the equilibrium bias condition.
In the experiments reported here, this is ensured by applying an ac signal at a power that is less than $1\%$ of the operating dc bias applied at frequencies $\omega_\mathrm{ac}$ ranging from $2\pi\times$10~Hz to $2\pi\times$30~kHz, with $\omega_\mathrm{ac}\leq 1/\tau_\mathrm{el}\ll1/\tau^\prime$.
This frequency range is chosen such that it covers both the thermal and electrical bandwidth of the detector.
From the high-frequency limit ($\sim 30 \, \mathrm{kHz} > 1/\tau$) we obtain an experimental estimation of $\beta_I$, as derived  in Eq.(10) of Ref.~\cite{taralli2019complex}.
This is consistent with the way we derive $\beta_I$ in the calculation shown below.
Note that the highest values for $\omega_\mathrm{ac}$, are orders of magnitude  smaller than the frequency of the Josephson oscillations.
This leads to a readout time $\tau_\mathrm{el}\gg\tau^\prime$, such that the voltage oscillations shown in Fig.~\ref{fig:device} are not captured.
An overview of the time constants involved in the dynamical system is given in Table~\ref{tab:Timescale}.

\begin{table}[h]
\begin{tabular}{c}
\end{tabular}
\begin{tabular}{|c | c | c | c |}
\hline
\cline{1-4}
\hline
Symbol &  Time scale  & Frequency scale  & Description \\
\cline{1-4}
\hline
$\tau$   &   $\gtrsim$10~ms & $\lesssim$100~Hz &  Thermal \\
$\tau_\mathrm{el}$  & $\lesssim$1~ms    &  $\gtrsim$ kHz  &  Electrical   \\
$\tau'$       & $\lesssim$100~ns    & $\gtrsim$10~MHz    & Josephson    \\
$\omega_\mathrm{ac}$       & 0.03 - 100 ms    & 0.01-30 kHz    & ac perturbation   \\
\hline
\cline{1-4}
\end{tabular}
\caption{Overview the time constants involved in the dynamical  system. The value of $\tau$ and $\tau'$ depends on the TES geometry and the normal resistance value $R_N$. The electrical bandwidth is typically defined  by the read-out circuit and the TES parameters.}
 \label{tab:Timescale}
\end{table}

Our approach to modeling the TES dynamical response derives from the observations we discuss above, namely that $V^\prime$ is reached at a timescale that is several orders of magnitude smaller than the total equilibration time ($\tau\gg\tau^\prime$) and that $\beta_I$ is extracted from the impedance measurement at the high-frequency limit $\sim 30 \, \mathrm{kHz} > 1/\tau$.
We may express the voltage response to a perturbing current, ${\delta I}$, as $V_0(I_0+{\delta I},T_0)=\eta V^\prime(I_0+{\delta I},T_0)$, where $\eta(I_0,T_0,G,n,R_N)\equiv V_0/V^\prime$ is the ratio of the equilibrium and Josephson voltages at bias current, $I_0$, and temperature, $T_0$.
Equivalently, $\eta$ can be defined as the ratio of the generated power at thermal equilibrium, $I_0V_0$, to the power generated by the electrical circuit, $I_0V^\prime$, in the absence of a thermal connection to the bath (see Fig.~\ref{fig:device}(a)).
From the definition of $\eta$ it follows that the TES dynamic resistance, $R_d\equiv\at{\frac{\partial V_0}{\partial I}}{T_0}$, can be written as
\begin{equation}
R_d=\eta \at{\frac{\partial V^\prime}{\partial I}}{T_0}.
\label{eq:Rd_def}
\end{equation}
Similarly, the voltage susceptibility to temperature perturbations within the dynamical response measurement is given by $\at{\frac{\partial V_0}{\partial T}}{I_0}=\eta \at{\frac{\partial V^\prime}{\partial T}}{I_0}$.
Therefore, the logarithmic temperature and current sensitivities at $I_0, T_0$ are given by, 
\begin{equation}
\alpha_I=\frac{T_0}{V^\prime}\at{\frac{\partial V^\prime}{\partial T}}{I_0},~\beta_I=\frac{R_d}{R_0}-1.
\label{eq:AlphaBeta_def}
\end{equation}

In order to make predictions for $\alpha_I$ and $\beta_I$ we first need to determine both the equilibrium voltage, $V_0(I_0,T_0)$, and the Josephson voltage, $V^\prime(I_0,T_0)$.
Due to the large timescale separation of $\tau$ and $\tau^\prime$ (typically $\tau/\tau^\prime>10^3$) solving both Eqs.~(\ref{eq:Fokker-Planck}) and (\ref{eq:PowerThermal}) up to electrothermal equilibrium would require a huge amount of resources.
In our approach, we require that, for each $I_0$, the equilibrium values $T_0$ and $V_0$ are simultaneous solutions of Eq.~(\ref{eq:Fokker-Planck}) at $t\rightarrow0$ and Eq.~(\ref{eq:PowerThermal}) at $t\rightarrow\infty$.
By doing this we essentially impose the thermal equilibrium solution, $V_0=K(T{}^n-T_0^n)/I_0$, to be the initial solution of the Fokker-Planck, \emph{i.e.}, $\langle V\rangle(t=0)\equiv V_0$.
Note that the steady state of Eq.~\ref{eq:Fokker-Planck} does not depend on the initial condition, \emph{i.e.}, for given $I_0$ and $T_0$ it yields the same $V^\prime(I_0,T_0)$ regardless of the initial value of $\langle V\rangle(t=0)$, however, solving them together and imposing $\langle V\rangle(t\rightarrow0)= V_0$ is necessary for obtaining the required equilibrium parameters $I_0,T_0$.

\begin{figure*}[t]
  \begin{center}
  \includegraphics[width=0.7\textwidth]{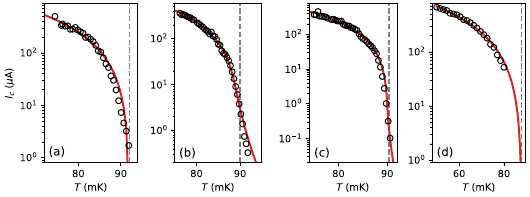}
  \end{center}
  \caption{
  Plots of the critical current $I_c$ as a function of temperature $T$ for the dc-biased devices: (a)~$40\cross60$, (b)~$50\cross50$, (c)~$70\cross50$ and (d)~$50\cross13$ geometry.
  The experimental data (open circles) is fitted with a combined Ginzburg-Landau and weak-link theory model (see main text for details).
 The dashed line indicates the critical temperature obtained from the fits.
 Note that for the $40\cross60$ and $50\cross13$ we only used the Ginzburg-Landau function for $I_c(T)$, due to lack of experimental data at higher temperatures.
  }
  \label{fig:Supp_Ic}
\end{figure*}

Furthermore, the simulation takes as input the fitted $I_c(T)$ curve, an example of which is depicted in Fig.~\ref{fig:device}(b), and the measured material- and geometry- dependent parameters $G, n, R_N$, as well as the bath temperature $T_b$, which are kept constant.
As input in our model, we additionally use the experimentally obtained device parameters summarized in Table~\ref{tab:Parameters}, together with the critical temperature, $T_c$, obtained from the $I_c(T)$ fit.
Fig.~\ref{fig:device}(b) shows $I_c(T)$ for the $80\cross10$ device.
The data and $I_c(T)$ fits for the other devices are demonstrated in Fig.~\ref{fig:Supp_Ic}.
For the device characterization we used the same methods as reported in Ref.~\cite{gottardi2021voltage}.

\begin{table}[h]
\begin{tabular}{c}
Parameters vs device geometry (Length$\cross$Width) 
\end{tabular}
\begin{tabular}{|l | c | c | c | c| c|}
\hline
\cline{1-6}
\hline
L$\cross$W ($\mu\mathrm{m}^2$) &  $40\cross60$ & $50\cross50$ & $70\cross50$ & $50\cross13$ & $80\cross10$ \\
\cline{1-5}
\hline
$R_N (\mathrm{m}\Omega)$   & 9.3      & 14.0 & 19.6   &53.8   & 206.4 \\
$T_b$~(mK)       &50    & 50    & 50    & 50  & 53   \\
$G$~(pW/K)       & 70     & 65    & 82  &43   & 60   \\
$n$              & 3.0     & 3.0   & 3.3  &3.0   & 3.3  \\
$C_H$~(pJ/K)     & 0.89     & 0.87    & 0.87  &0.84   & 0.78    \\
$T_c$~(mK)       & $92.5$     & $90.0$   & $90.3$  &$87.8$  & $81.4$   \\
\hline
\cline{1-6}
\end{tabular}
\caption{Table of device parameters.}
 \label{tab:Parameters}
\end{table}

\begin{figure}[t]
  \begin{center}
  \includegraphics[width=1\linewidth]{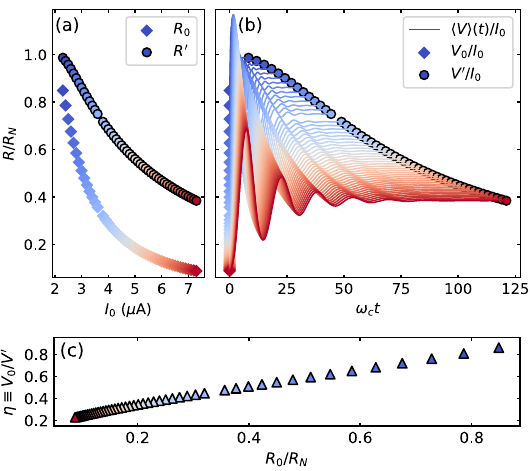}
  \end{center}
  \caption{
  {(a)~Normalized equilibrium resistance $R_0/R_N$ (diamonds) and Josephson resistance $V^\prime/(I_0R_N)$ (circles) as a function of bias current.
  (b)~Transient dynamics of $\langle V\rangle(t)$.
  For each set of $I_0, T_0$ we solve the Fokker-Planck equation with the initial and final states corresponding to $V_0/(I_0R_N)$ (diamonds) and $V^\prime/(I_0R_N)$ (circles).
  Note that for each set the simulation stops at $t=8T_J$, where $T_J$ is the oscillation period, $T_J=2\pi/(\omega_c \sqrt{(I_0/I_c)^2-1})$.
  (c)~Corresponding ratio $\eta \equiv V_0/V^\prime$ as a function of $R_0/R_N$.
  The color coding is the same in all figures, with blue and red indicating the lowest and highest bias currents, respectively.
   }
  }
  \label{fig:R0vsFP}
\end{figure}

Having obtained $V_0$, for each $I_0,T_0$, we then determine the Josephson voltage, $V^\prime$, by numerically solving the Fokker-Planck equation with the initial condition ${\langle V\rangle(t=0)}=V_0$, not including any thermal dynamics.
As an example, in Fig.~\ref{fig:R0vsFP}(a) we plot the steady-state resistance, $R_0=V_0(I_0,T_0)/I_0$, (diamonds) and the associated \emph{Josephson resistance}, $V^\prime(I_0,T_0)/I_0$, (circles) as a function of the current $I_0$.
Each set of $I_0$ and $T_0$ is represented with a different color, from low (blue) to high (red) current.
The computation is performed using the $I_c(T)$ curve shown in Fig.~\ref{fig:device}(b) and parameters $R_N=206.4~\mathrm{m}\Omega$, $G=50~\mathrm{pW/K}$, $n=3.3$, $T_b=50~\mathrm{mK}$.

\begin{figure*}[t]
  \begin{center}
  \includegraphics[width=0.9\textwidth]{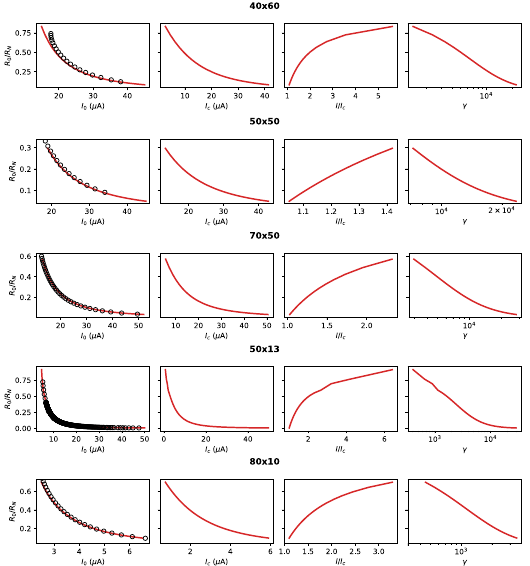}
  \end{center}
  \caption{
  The first column from the left shows the equilibrium normalized resistance $R_0/R_N$ as a function of current for each device.
  The red curve is the simulation result, $V_0/(I_0R_N)$, while the open circles indicate the experimentally obtained resistances for the corresponding device.
  In addition, for each device we include, from left to right, the numerically obtained values for $I_c(T_0)$, $I_0/I_c(T_0)$ and $\gamma(T_0)$ across the resistance curve.
  These equilibrium values are used as input to the Fokker-Planck equation when calculating $V^\prime(I_0,T_0)$ and determining $R_d$, $\alpha_I$ and $\beta_I$.
  }
  \label{fig:Supp1}
\end{figure*}

Furthermore, in Fig.~\ref{fig:R0vsFP}(b), we plot the time evolution of the \emph{Josephson resistance}, ${\langle V\rangle(t)}/I_0$, with initial and final values corresponding to $R_0(I_0,T_0)$ and $V^\prime(I_0,T_0)/I_0$, respectively, with the same color coding as in Fig.~\ref{fig:R0vsFP}(a).
In addition, in Fig.~\ref{fig:R0vsFP}(c) we plot $\eta=V_0/V^\prime$ as a function of the resistance $R_0$.
Note that $\eta<1$ is expected since, due to thermal losses, the power at thermal equilibrium, $I_0V_0$, is always lower than the power $I_0V^\prime$ that the circuit generates in the absence of thermal coupling to the bath.
In Fig.~\ref{fig:Supp1}, for each device, we plot our numerical results of $R_0$ as a function of $I_0$, as well as $I_c(T_0)$ and $\gamma(T_0)$.
Our theoretical predictions for $R_0$ are in excellent agreement with the experimentally obtained resistances at thermal equilibrium.

Having obtained both voltages  $V_0(I_0,T_0)$ and $V^\prime(I_0,T_0)$, we proceed to calculate the dynamic resistance, $R_d$, as well as $\alpha_I$ and $\beta_I$, as defined in Eq.~(\ref{eq:AlphaBeta_def}).
More specifically, in order to obtain $R_d$ and $\beta_I$, we additionally solve the Fokker-Planck equation at $I_0+\delta I$, for infinitesimally small $\delta I$, and determine the partial derivative $\at{\partial V^\prime/\partial I}{T_0}=\frac{V^\prime(I_0+\delta I,T_0)-V^\prime(I_0,T_0)}{\delta I}$.
Similarly, we find $\at{\partial V^\prime/\partial T}{I_0}$ by calculating $V^\prime(I_0, T_0+\delta T)$ for $\delta T\rightarrow0$, and then determine $\alpha_I$ following Eq.~(\ref{eq:AlphaBeta_def}).

\begin{figure*}[t]
  \begin{center}
  \includegraphics[width=1\textwidth]{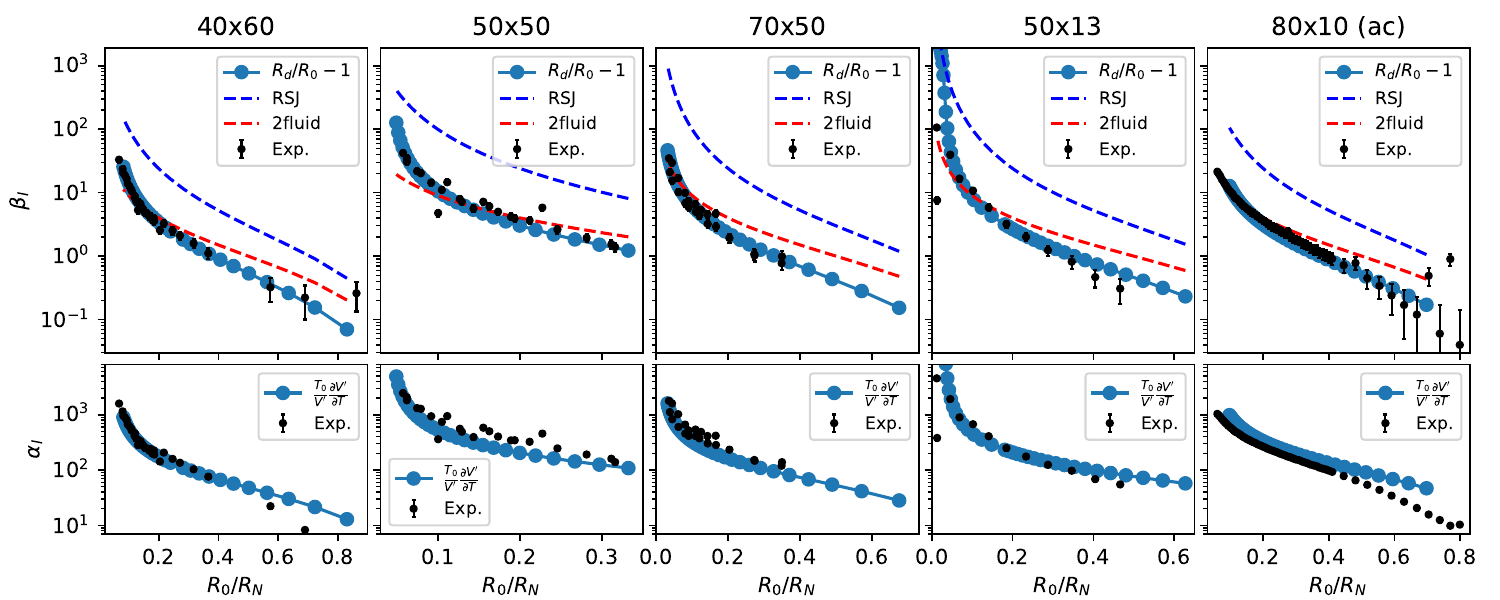}
  \end{center}
  \caption{
  {Logarithmic current and temperature sensitivities, $\beta_I$ (top) and $\alpha_I$ (bottom), respectively, vs $R_0/R_N$ for different device geometries, shown from low ($40\cross60$) to high aspect ratio ($80\cross10$).
  In the experiment, the first four devices from the left are dc-biased while the last one is ac-biased at $1.1$~MHz.
  The predictions using our model are plotted in cyan-blue circles, showing excellent agreement with the data (black).
  Less agreement is obtained for the ac-biased $80\cross10$ pixel, possibly due to the fact that our model assumes a dc source.
  For comparison we plot the theoretical predictions based on the pure RSJ model (blue dashed) and the two-fluid model with $c_R=1$ (red dashed).
  }
  }
  \label{fig:BetaAlphaVSR0}
\end{figure*}

\section{Results}
Our numerically calculated predictions for $\alpha_I$ and $\beta_I$ are plotted in Fig.~\ref{fig:BetaAlphaVSR0} as a function of $R_0$ (cyan-blue circles) for several TES bilayer geometries, showing excellent agreement with the measured values (black dots).
For comparison, we also plot the analytical predictions resulting from the RSJ model in the absence of thermal effects, $\beta_I^\mathrm{RSJ}=(R_N/R_0)^2-1$ (blue dashed), and the two-fluid model, $\beta_I^\mathrm{2fluid}=c_RR_N/R_0-1$ (red dashed) with $c_R=1$~\cite{bennett2013resistance}.
The latter is based on the assumption that the TES resistance is determined, instead of Josephson phenomena, by other physical processes occurring in the thin film, such as phase-slip events~\cite{irwin1998thermal}.
The experimental data (black circles) is obtained via complex impedance measurements~\cite{taralli2019complex,lindeman2004complex} using the same measurement setup that is reported in Ref.~\cite{gottardi2021voltage} both for the dc- and ac-biased devices.

\begin{figure*}[t]
  \begin{center}
  \includegraphics[width=1\textwidth]{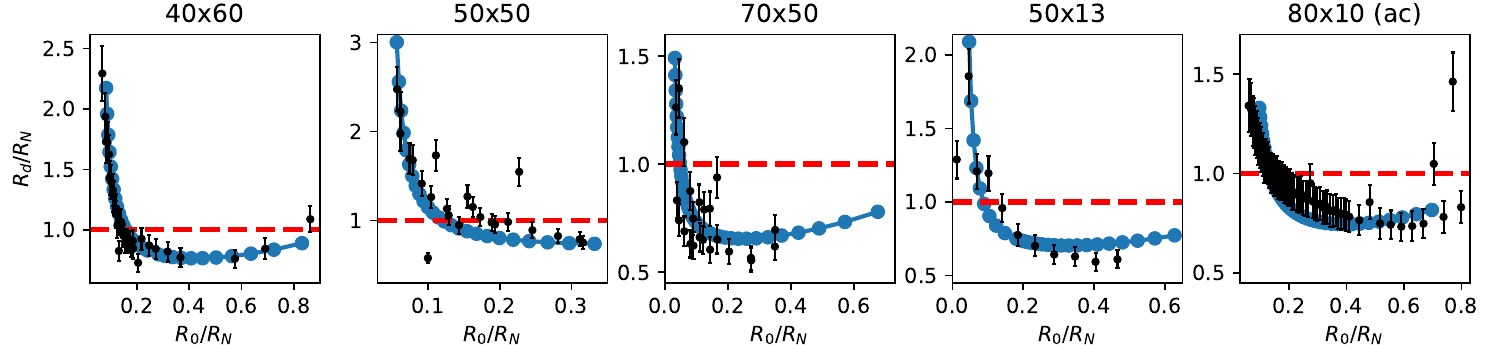}
  \end{center}
  \caption{
  {Dynamic resistance vs $R_0/R_N$. The predictions based on our model are plotted in cyan-blue circles, showing excellent agreement with the data (black). The dashed horizontal line indicates the prediction of the two-fluid model with $c_R=1$.
  }
  }
  \label{fig:RdVSR0}
\end{figure*}

Evidently, $\beta_I^\mathrm{RSJ}$ overpredicts the data over the whole range of the transition, signaling that pure Josephson physics is not enough to explain the TES response.
Note that even when thermal noise is included, \emph{e.g.}, using the Fokker-Planck formulation of Eq.~(\ref{eq:Fokker-Planck}) or analytical calculations~\cite{kozorezov2011modelling}, this trend persists.
In addition, comparative analysis has shown that the two-fluid model predictions are much closer to the data, therefore, challenging the idea of using weak-link models to describe TES physics~\cite{bennett2013resistance}.
However, as Fig.~\ref{fig:BetaAlphaVSR0} demonstrates, our weak-link model that additionally includes the effects of the power-thermal equation in the parameter $\eta$, leading to the updated definition of $R_d$ in Eq.~(\ref{eq:Rd_def}), predicts the correct behavior for both $\alpha_I$ and $\beta_I$.
The predictions for $\alpha_I$ based on the RSJ and two-fluid model are not shown, since, as pointed out in Ref.~\cite{bennett2013resistance}, a fair comparison of the two models is hindered by a strong dependence on the derivative $\partial I_c/\partial T$.

While the two-fluid model prediction is also close to the experimental data, the situation changes drastically when looking at the dynamic resistance behavior.
More specifically, the two-fluid model prediction for the dynamic resistance is $R_d^\mathrm{2fluid}=c_RR_N$, where $c_R$ is a phenomenological parameter~\cite{ullom2015review}.
In the standard version of the model $c_R$ is constant~\cite{bennett2013resistance}, leading to a constant dynamic resistance, which is in stark contrast with our observations.
In particular, as shown in Fig.~\ref{fig:RdVSR0}, our experimental data shows that $R_d$ varies significantly across the resistance curve, with a steep increase at low bias $R_0/R_N\lesssim0.2$, which is in excellent agreement with our model predictions (cyan-blue).

Alternative mechanisms for this behavior could be the formation of different numbers of phase-slip lines forming across the resistance curve, which could yield a varying two-fluid parameter $c_R$~\cite{bennett2013resistance,bennett2014phase}.
However, due to the large complexity of this model, a consistent model prediction for $c_R(R_0)$ remains elusive~\cite{bennett2014phase} and $c_R$ is typically only used as a fit parameter~\cite{morgan2017dependence}.
On the other hand, as we show, the correct behavior of $R_d(R_0)$ can be directly predicted from physical mechanisms arising from the Josephson effect.
Our results, together with recent observations of a nonlinear complex impedance in TES~\cite{gottardi2021voltage}, provide strong evidence that weak-link physics play a crucial role in the dynamical response of TES.

\begin{figure}[t]
  \begin{center}
  \includegraphics[width=\linewidth]{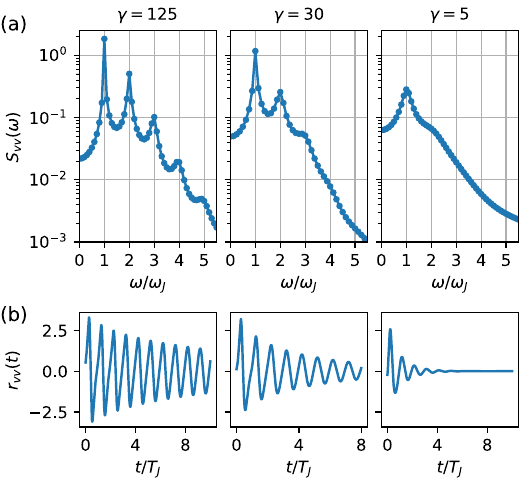}
  \end{center}
  \caption{
  {(a)~Power spectral density calculated for several values of $\gamma$ at fixed bias $I_0=2I_c$, using the parameters of the $80\cross10$ device.
  The peaks correspond to integer multiples of the Josephson frequency $\omega_J\equiv\omega_c\sqrt{(I_0/I_c)^2-1}$.
  As the temperature increases (decreasing $\gamma$) less peaks are resolved, due to larger oscillation damping, and $S_{vv}(0)$ increases.
  The corresponding autocorrelation functions $r_{vv}(t)$ for each plot are shown in (b) in units of the oscillation period $T_J$.
  }
  }
  \label{fig:Corr}
\end{figure}

Our approach not only leads to accurate predictions for the TES dynamical response, but also provides a powerful tool for modeling noise in TES.
More specifically, via the Fokker-Planck equation we can calculate the autocorrelation function $r_{vv}(t-t^\prime)=(\langle V(t) V(t^\prime)\rangle-\langle  V(t)\rangle \langle V(t^\prime)\rangle)/(I_cR_N)^2$.
Following the Wiener–Khinchin theorem, when $r_{vv}$ is Fourier-transformed, it yields the power spectral density, $S_{vv}(\omega)=\int r_{vv}e^{-i\omega t }dt$~\cite{kogan2008electronic}.
This is one of the most important quantities in a random process since it carries information regarding the contribution to noise from all frequencies.
As a proof of principle, in Fig.~\ref{fig:Corr}(a) and \ref{fig:Corr}(b) we plot the calculated $S_{vv}$ and $r_{vv}$, respectively, for the $80\cross10$ device at bias $I_0=2I_c$ for three different values of $\gamma$.
As expected, in Fig.~\ref{fig:Corr}(a) we see well-defined peaks at frequencies equal to the Josephson frequency and its higher harmonics.
For smaller $\gamma$, \emph{i.e.}, stronger thermal noise damping, the peaks at higher harmonics start to disappear and the spectral density gets broader resulting in more noise at $\omega\rightarrow0$.

With this method one can, therefore, estimate the entire spectral density numerically, including contributions to noise from all the harmonics.
Importantly, however, for an accurate estimation of $S_{vv}$ one would require increased computational resources as the simulation time and number of data points necessary to properly resolve the spectrum need to increase with increasing $\gamma$.
Overcoming these issues may therefore enable a quantitative analysis as well as a direct comparison of our experimental data ($\gamma\sim10^3$) and analytical calculations of $S_{vv}(0)$~\cite{kogan1988fluctuation,nagaev2019josephson,wessels2021model}.
In particular, existing analytical expressions of spectral densities in the RSJ model take into account processes up to the first Josephson harmonic~\cite{likharev1972fluctuation,gottardi2021voltage}.
Our method could therefore extend the current predicting capabilities beyond this approximation since it inherently takes into account all higher Josephson harmonics.
Further work in this direction may help answering one of the most important open questions in the field, \emph{i.e.}, to determine the extent to which the mixed-down high-frequency noise contributes to the observed noise at low frequencies.
A thorough investigation of this process could shed further light into the mystery of measured ``excess noise'' in TES~\cite{wessels2021model,gottardi2021voltage}.

\section{Conclusion}

We have found strong evidence of weak-link physics in the dynamical response of TES.
We have developed a theoretical framework, using a Fokker-Planck description, that enables the incorporation of both Josephson and electrothermal effects, and accurately reproduces the observed noise sensitivity parameters while outperforming existing models.
In particular, we theoretically predict and experimentally observe a varying dynamic resistance across the TES resistance curve, which is a key signature of Josephson physics and rules out the standard two-fluid model prediction with fixed $c_R$.
Furthermore, our model allows us to compute the power spectral density directly, enabling a further investigation on the role of mixed-down high-frequency noise that is believed to contribute to unexplained ``excess noise'' in TES~\cite{wessels2021model,gottardi2021voltage}.
Our description can also be extended to a 2D Fokker-Planck formulation, offering the opportunity to model TES structures with additional capacitance~\cite{koch1980quantum}.
Finally, our theoretical framework could be employed to model structures obeying non-sinusoidal current-phase relations~\cite{golubov2004current}, such as long diffusive SNS junctions~\cite{lhotel2007divergence}.

\section{Acknowledgments}
We thank J.-R. Gao for fruitful discussions.
This work is funded by the European Space Agency (ESA) under ESA CTP Contract No. 4000130346/20/NL/BW/os.

\bibliography{../../BibMarios}

\end{document}